\documentclass[sn-mathphys-num,iicol]{sn-jnl}% Math and Physical Sciences Numbered Reference Style 

\usepackage{breqn}
\usepackage{graphicx}%
\usepackage{multirow}%
\usepackage{amsthm}%
\usepackage{amsfonts}
\usepackage{mathrsfs}%
\usepackage[title]{appendix}%
\usepackage{xcolor}%
\usepackage{textcomp}%
\usepackage{manyfoot}%!TEX encoding = UTF-8 Unicode
\usepackage{booktabs}%
\usepackage{algorithm}%
\usepackage{algorithmicx}%
\usepackage{algpseudocode}%
\usepackage{listings}%
\usepackage{float}
\begin{document}
\title{Life Cycle Analysis for Emissions of Scientific Computing Centres}

\author[1]{\fnm{Mattias} \sur{Wadenstein}}\email{maswan@ndgf.org}

\author*[2]{\fnm{Wim} \sur{Vanderbauwhede}}\email{wim.vanderbauwhede@glasgow.ac.uk}

\affil[1]{\orgdiv{HPC2N}, \orgname{Umeå University}, \orgaddress{\city{Umeå}, \country{Sweden}}}

\affil*[2]{\orgdiv{School of Computing Science}, \orgname{University of Glasgow}, \orgaddress{ \city{Glasgow} \country{UK}}}

\abstract{
We propose a dedicated model to assist with the life cycle analysis of emissions
of scientific computing centres. The model takes into account both the embodied carbon and emissions from use, as well as other factors such as data centre power usage efficiency, data centre expansion, hardware replacement, increase in energy efficiency of next-generation hardware, reduction in carbon intensity of the
electricity supply and potential for heat reuse. If differs from existing models in its detailed handling of hardware embodied carbon and time dependency of various factors affecting the emissions. We present a number of scenarios where we apply the model to real-life HPC centres in different countries to illustrate how the trade-offs depend on the various factors and validate our model against the literature.
}


\maketitle
\section{Introduction and research questions}

Emissions from scientific computing are considerable, and in view
of the climate emergency, HPC centre operators and users should be
aware of this and work towards minimising the emissions. Both the
practice of scientific code development and deployment, as discussed
e.g. in our report \cite{wim_vanderbauwhede_2023_7709483}, and the HPC centre operation and procurement,
discussed in \cite{10.1145/3706598.3713919}, play an important role in this process. In
this paper, we specifically look at life cycle analysis of emissions
of scientific computing centres, to allow HPC centre operators to
minimise the overall emissions of the workloads running on the infrastructure. 

The questions we set out to answer are:
\begin{itemize}
\item What is the contribution of server embodied carbon, infrastructure embodied carbon, electricity generation carbon intensity and facility expansion on the cumulative emissions of an HPC facility?
\item What is the optimal replacement cycle for scientific computing hardware,
from a total CO$_2$e emissions point of view?
\end{itemize}

To answer these questions, we have developed a model to assist us
in making the carbon lifecycle analysis of a scientific computing
centre. The model equations apply more generally to large data centres,
but some of our assumptions are specific to scientific computing.

\section{Related Work}

There are a number of publications on LCA for data centres \cite{Ma2024LCA,whitehead2015LCA,samaye2025LCA,schneider2023}, although none specifically for HPC facilities. \cite{schneider2023} uses an in-house developed model using SAP Xcelsius; \cite{whitehead2015LCA} uses the commercial package SimaPro; \cite{samaye2025LCA} uses an in-house model with inputs for embodied carbon from the NegaOctet database; \cite{Ma2024LCA} is cost-based and does not provide details on the input data beyond the costs. The main differences and contributions of our work compared to this related work are:

\begin{itemize}
\item The focus on overall CO$_2$ emissions for HPC facilities, which manifests itself in the parameters of the facilities and the hardware and in the use of the HEPScore benchmark as normalisation metric.
\item The inclusion of a detailed model for server embodied carbon. The existing work relies instead on existing databases.
\item The focus in the existing work is much more on the physical infrastructure than the IT equipment.
\item The model for expansion and replacement and other explicit time dependencies as expressed in our work is not present explicitly in any of the related work. 
\item None of the related work provides access to the model source code.
\end{itemize}

\section{Methodology}

Models for life cycle analysis are quite common. The non-commercial ones are frequently implemented using spreadsheets (even the model from Schneider Electric \cite{schneider2023} is based on spreadsheets). For more complex models, this is not practical, as the equations become complex and are often recursive. Furthermore, spreadsheets are notoriously error-prone. We have therefore implemented our model in the functional programming language Haskell, which allows great flexibility in the expression of mathematical equations and reduces the scope for errors. The code can be found in \cite{hpc_lca_code_wv2025}. It consists of a LCA model, discussed in the next section, and a model for the calculation of the embodied carbon of the hardware, discussed in Section \ref{subsec:Embodied-carbon-model}.

\subsection{LCA model}

The total carbon footprint for a given capacity over time \emph{t}
is the sum of the embodied carbon of the hardware (including manufacturing,
transport and end-of-life disposal) and the runtime emissions (Eq. \ref{eq1}). In all following equations, $t \in \mathbb{N}$.
\begin{eqnarray}
\mathtt{total\_emissions}(t) ~ = & ~ \\
& \mathtt{runtime\_emissions}(t)\nonumber \\
& +\mathtt{embodied\_emissions}(t)\nonumber 
\label{eq1}
\end{eqnarray}

\subsubsection{Use of HEPScore 2023 benchmark}\label{subsubsec:HS23}

Rather than expressing the compute power of our nodes in cores FLOPS, we express them instead in terms of the HEPScore23 (HS23) benchmark.
HEPScore23 \cite{szczepanek2024hep} is a benchmark based on high-energy physics (HEP) applications created by the HEPiX Benchmarking Working Group on behalf of the Worldwide LHC Computing Grid (WLCG). WLCG coordinates the computing and networking resources on behalf of the experiments at the Large Hadron Collider at CERN. This allows comparison of different HPC centres by expressing the performance with a single numerical value that correlates closely to how much LHC scientific work can be done per time unit.

The values are in the 20-40 range per core, which means that whole compute nodes or sites
are often in the WLCG contexts referred to with the SI k and M
prefixes for kilo- and mega-, in this paper we use kHS23, which is also the form we
get the benchmark numbers for hardware we use as input.

For other scientific workloads, other benchmarks may be more accurate. However, using the HS23 values we can compare the different facilities despite their different configurations. Especially for our yearly cluster expansion scenarios we feel it important to express the size requirements in terms of useful compute capability.

\subsubsection{Site parameters}

The emissions depend on the site in various ways, as discussed
in Section \ref{subsec:Assumptions}. The factors taken into account are:

\begin{itemize}
    \item yearly site expansion factor, $\alpha$
    \item yearly average site load factor, $\lambda=0.8$    
    \item site power usage effectiveness, $\textit{PUE}$
    \item site explicit heating/cooling emissions, $\textit{ECE}$ (kgCO$_2$e/kWh) 
    \item site node lifetime, $\Delta t$ (years) 
    \item \#nodes per kHS23, $n_{kHS23}$
    \item node embodied carbon, node$_{ec}$ (kgCO$_2$e) 
    \item energy consumption per kHS23 per year, $E_{kHS23}$ (kWh/year)
    \item node idle power consumption factor, $\gamma=0.3$
    \item location electricity~carbon~intensity $CI$ (kgCO$_2$e/kWh) 
\end{itemize}

The energy consumption per kHS23 per year, $E_{kHS23}$, takes into account the load and the idle power consumption:
\begin{equation}	\label{eq1b}
	E_{node} = (\lambda+(1-\lambda )\cdot\gamma)\cdot E_{kHS23}/n_{kHS23}
\end{equation}

The node embodied carbon is discussed in detail in Section \ref{subsec:Embodied-carbon-model}. The site explicit heating/cooling emissions $\textit{ECE}$ are explained in Section \ref{subsubsec:facility-cooling}.

\subsubsection{Data centre expansion and server replacement}

In practice, the HPC centre will periodically renew its hardware
and also expand its capacity. This has an effect on both the embodied
and runtime emissions: on the one hand, replacing hardware increases
the embodied carbon; on the other hand, newer generation hardware
is more energy efficient, leading to reductions in runtime emissions.

We model the expansion as

\begin{equation}\label{eq2}
\mathtt{expansion}(t,\alpha) = (1+\alpha)^{t - 1} 
\end{equation}
~\\
The replacement is conceptually modelled using a generic step function:

\begin{equation}\label{eq3}
\mathtt{step}(t,\Delta t) =  \Delta t \cdot \left\lfloor \frac{t-1}{\Delta t}\right\rfloor + 1 
\end{equation}

\subsubsection{Runtime emissions}

The runtime emissions depend in first order on the electricity
consumption of the hardware, the power usage effectiveness (PUE) of the
data centre and the electricity carbon intensity (amount of CO$_2$ emitted
per unit of electricity produced) of the electricity generation.

We assume (see \ref{subsec:Assumptions})
that energy efficiency follows Koomey's law \cite{koomey2011web}, which expresses and exponential
increase in efficiency over time:
\begin{eqnarray}\label{eq4}
\mathtt{compute\_efficiency\_correction}(t,\beta)= \nonumber\\ 
 \left(\frac{1}{1+\beta}\right)^{t-1} 
\end{eqnarray}

We use a value of $\beta\,=\,0.17$ based on data from Masanet et al. \cite{doi:10.1126/science.aba3758}, who show that the relative change in data centre energy use between 2010 and 2018 is $0.24$.

Furthermore, carbon intensity should improve over time and depends on the
geographical area where the centre is located. 

The function expressing this is (Eq. \ref{eq5}):
\begin{eqnarray}\label{eq5}
\mathtt{carbon\_intensity\_correction}(t,\kappa)= \\ 
(1-\kappa ) ^{t-1}\nonumber 
\end{eqnarray}

So that
\begin{eqnarray}\label{eq6}
\mathtt{carbon\_intensity}(t,\kappa) & = &  ~ \\ 
CI \cdot \mathtt{carbon\_intensity\_correction}(t,\kappa)\nonumber & ~ & ~
\end{eqnarray}

For example, based on data from \cite{OWiD2024a}, and extrapolating the current trends over the past decade, for Sweden it would be $\kappa = 0.01$; for the UK it would be $\kappa = 0.1$, and for the global electricity carbon intensity it would be $\kappa = 0.025$. 

With these assumptions, the expression for runtime emissions per year and kHS23 becomes (Eq. \ref{eq7}):
\begin{eqnarray}\label{eq7}
\mathtt{runtime\_emissions\_year}(t) = ~ ~\\
    E\_{kHS23} \nonumber \\
 \cdot(\mathtt{carbon\_intensity}(t,\kappa)+\textit{ECE}) \cdot \textit{PUE}   \nonumber \\
 \cdot\mathtt{compute\_efficiency\_correction}(step(t,\Delta t),\beta) \nonumber\\
 \cdot\mathtt{expansion}(t,\alpha) \nonumber
\end{eqnarray}

Cumulative emissions from electricity consumption are obtained by summing the yearly emissions (Eq. \ref{eq7b}):
\begin{eqnarray}\label{eq7b}
\mathtt{runtime\_emissions}(t) ~=~ \\
\sum_{t_c=1}^{t} \mathtt{runtime\_emissions\_year}(t_c) \nonumber
\end{eqnarray}

\subsubsection{Embodied emissions}

For the embodied emissions of the site, we accumulate the embodied carbon of the new replacement cycle to the total embodied carbon at the time. As we replace all servers after a lifetime $\Delta t$, and we have a yearly cluster expansion, we get a combination between a stepwise replacement and the gradual replacement due to yearly expansion. The resulting expression for the embodied emissions over time is a recursive function. The model is constructed as follows:

\begin{itemize}
\item $n_{inst}$ is the number of nodes installed in a given year. 
This is the initial number multiplied with the expansion factor year by year, in other words the expansion is exponential.
\begin{equation}
\begin{aligned}
n_{inst}(t)~=~
	n_{kHS23} \cdot 
	(1+\alpha)^t
\end{aligned} \label{eq10a}
 \end{equation}
 ~\\
\item $n_{new}$ is the new servers installed in a given year. 
This is made up of the additional nodes due to the yearly cluster expansion plus the retired nodes that have to be replaced.
\begin{equation}
\begin{aligned}
n_{new}(t,\Delta t)~=~\begin{cases}
n_{kHS23} & ,~t=0\\
n_{inst}(t)+n_{ret}(t,\Delta t) & ,~t>0
\end{cases}
\end{aligned} \label{eq10b}
 \end{equation}
  ~\\
\item $n_{ret}$ is the number of retired server nodes in a given year. This is equal to the number of server nodes installed $\Delta t$ years before.
 \begin{equation}
\begin{aligned}
n_{ret}(t,\Delta t) & =\begin{cases}
0 & ,t<\Delta t\\
n_{new}(t-\Delta t,\Delta t) & ,t\geq\Delta t
\end{cases}
\end{aligned} \label{eq10c}
 \end{equation}
~\\
\end{itemize}
The equation for the embodied emissions is shown in Eq. \ref{eq10}. The embodied carbon corresponds to the new nodes, so we have

\begin{eqnarray}\label{eq10}
\mathtt{embodied\_emissions\_year}( t, \Delta t , node_{ec})~=~\\
node_{ec} \nonumber\\
 \cdot\, n_{new}(t,\alpha,\Delta t) \nonumber\\ 
 \cdot\, \mathtt{embodied\_carbon\_correction}(t)\nonumber
\end{eqnarray} 

The reduction in embodied carbon with every new hardware generation, $\mathtt{embodied\_carbon\_correction}(t)$, is modelled as:

\begin{equation}
\mathtt{embodied\_carbon\_correction(t)} = (1-\epsilon)^{t-1} \label{eq8}
\end{equation}
~\\
This not site-dependent. The factor $\epsilon$ in Eq. \ref{eq8} is currently not known: for carbon intensity, there are projections; for embodied carbon, there is no data, but the emissions from the electricity used for manufacturing follow the same trend. We used a value of $\epsilon=0.00265$ which corresponds to a reduction of 5\% over 20 years. This is likely to be conservative as recycling of the hardware would also effectively lower the emissions; for a discussion of the rationale, see Section \ref{Sec:embodied-carbon-trend}
%The assumptions regarding the correction of embodied carbon may be very conservative. Since recycled materials and improved production processes can also lower embodied carbon, it would be valuable to briefly elaborate on these factors and their potential impact on the model.
%Discuss this under SSD

The final cumulative embodied carbon model (Eq. \ref{eqCumEmbEm}) includes two more factors to account for the data centre infrastructure, $node_{ec,if}$ which is the per-node portion of the embodied carbon of the construction of the data centre, and $\Delta node_{ec,if}$ which is the per-node portion of the yearly additional embodied carbon arising from maintenance of the infrastructure. The values are based on \cite{schneider2023}.

\begin{eqnarray}\label{eqCumEmbEm}
\mathtt{embodied\_emissions}(t) ~=~ \\
\sum_{t_c=1}^{t} ( \mathtt{embodied\_emissions\_year}(t_c) \nonumber\\
+ ~\Delta node_{ec,if} \cdot t_c ~) \nonumber\\
+ ~node_{ec,if} \nonumber
\end{eqnarray}

\subsection{Embodied carbon model}\label{subsec:Embodied-carbon-model}

The factor $n_{ec}$ for the contribution of the embodied carbon of the hardware in Eq. \ref{eq10} needs to be calculated separately. 
The data available for embedded carbon in a compute node that comes
from manufacturing, transport, and end-of-life disposal is available for a few
select server models from hardware vendor reports. None of these servers are
typical compute nodes for scientific computing.

To compute the embodied carbon of server manufacturing, we therefore re-implemented the model by Boavizta \cite{lorenzini2021digital} in Haskell. This model is very comprehensive but as it was published in 2021 and has not been updated since, we looked for more up-to-date estimates for various parameters. Beside estimates for the chips, the model also includes contributions from packaging, power supply, server enclosure and rack enclosure. 

In particular for the various chips used in the server (CPU, RAM, SSD) we use the ACT methodology (Architectural Carbon modelling Tool) \cite{10.1145/3470496.3527408}. This tool uses the electricity consumption of the manufacturing process, the embodied carbon for the materials, and the greenhouse gas potential for the various gases used in production, and combines this with the die size to obtain an estimate for the embodied carbon of the chip. We updated some of the parameters (those that were included without reference in the ACT paper) using data from \cite{10.1145/3630614.3630616}, \cite{9372004} and \cite{AnandTech2019}. We also extended the model to include non-integrated GPUs.

The model is based on the die size of the various chips in a server (CPU, GPU, RAM, SSD). The embodied emissions are calculated using process-specific data (embodied emissions for the materials and gases used, amount of electricity used, process yield) and the CI of the electricity used by the fab. For example, for the CPU we have Eq. \ref{eqEmbCarbonCPU}. The first line are the emissions from production of the die, which consist of emissions from the electricity used $en_p*ci_{fab}$, embodied emissions from the materials used in production $ec_m$ and embodied as well as incurred emissions from the gases used in production $ec_g$ (many of which have a much higher GWP than CO$_2$), corrected by the wafer yield. The second line takes into account the number of cores $n_{\textit{cores}}$ and units $n_{\textit{units}}$ in the CPU packages, as well as the embodied carbon of the base $ec_{\textit{base}}$ and of the package $( ( n_{\textit{cores}} \cdot s_{\textit{die}} + oh_p ) \cdot ec_p$ including the packaging overhead $oh_p$.
\begin{eqnarray}
        ec_p & = & (en_p*ci_{fab} + ec_m + ec_g)/\textit{yield} \\
        ec_{cpu} &= & n_{\textit{units}} \cdot ( ( n_{\textit{cores}} \cdot s_{\textit{die}} + oh_p ) \cdot ec_p + ec_{\textit{base}} )\nonumber
\label{eqEmbCarbonCPU}
\end{eqnarray}

Similar equations are used for RAM, SSD and GPU. The model further includes the embodied carbon of the power supplies, motherboard and enclosure and emissions from assembly.  The model also takes into account the embodied emissions of the facility and of the networking infrastructure, based on estimates worked out in \cite{schneider2023}: facility infrastructure embodied carbon is estimated at 1,829 tCO$_2$e on construction, increasing to 2,400 tCO$_2$e after 20 years due to maintenance and replacement of parts of the infrastructure, for a 1MW data centre; networking equipment embodied carbon is estimated at 6\% of the total server embodied carbon. We relate this through the individual nodes by scaling with the power consumption of the node, corrected by the data centre PUE. For full details, we refer to the source code and model documentation \cite{hpc_lca_code_wv2025}.

\subsection{Assumptions\label{subsec:Assumptions}}

\subsubsection{Compute node assumptions}

We assume the computational capabilities and power usage of the current (spring 2024) 
best in class compute node, a configuration of  2x AMD EPYC 9754 HT as per \cite{Britton2023}, with 1TB RAM and 0.5TB NVME SSD. Idle power consumption is based on measurements reported in \cite{amdepyc2019,speccpu2020}.

As discussed in Section \ref{subsubsec:HS23}, we normalise on a benchmark for scientific computing based on the workloads from the physics experiments Large Hadron Collider (HEPScore23). By doing so we can for a given scientific computing capability estimate the embedded carbon from the compute nodes needed as well as the power usage per year of running these.

The vendor reports on carbon emissions from servers also include assumptions of runtime
emissions, but we disregard these and rely on our own calculations because scientific computing has a much different load pattern \cite{cernITsust} and we have more accurate numbers from facility managers at several institutes\footnote{personal communications}.

\subsubsection{Marginal power consumption}

Calculating the addition 1 kWh of power consumption in a particular area can be made
in several different ways. If the facility is buying ``green'' power, looking at the
emissions as per contract is probably the lowest emissions. On the other hand, an
increase in power usage in a region will either lead to an increase of generation or
a change in import/export power.

With detailed knowledge of the power generation structure within a region, we can
make some estimates on what increased generation would mean (short term it is most
likely higher use of the power generation that is currently used for balancing), or
impact of decreased exports.

In addition to these, we can also assume that over a longer time the power generation
impact for our additional scientific computing load will be similar to the average
generation mix in that electricity region. This is the approach used in this paper
because we feel it is the fairest way of measuring the impact, and it does not require
special knowledge or complicated constructs (e.g. ``assume 20 percent of the 
additional power used in North Central Sweden would otherwise have been exported to Denmark, another 10
percent to Poland, ...'').

The simulation tool provided can easily be adjusted to reflect other marginal power
consumption assumptions by doing the calculations and providing the 
yearly average electricity generation carbon intensity as input.

\subsubsection{Facility cooling\label{subsubsec:facility-cooling}}

In the current model, we assume that the PUE remains constant for
the entire lifetime of the centre. It is possible to incorporate stepwise
changes in PUE but those would only result in relatively small corrections,
so we use an average PUE estimate.

As an alternative to PUE, we use explicit CO$_2$e emissions when the facility
cooling is not just based on compressors and fans powered by electricity.
This is the case for one of the scenarios for site HPC2N, where district cooling is used
instead.

\subsubsection{Scenarios}
The scenario modelled is a site that has decided to contribute to scientific
computing by providing compute resources to researchers for the next 20 years,
starting with an empty room. We measure the scientific computing capacity 
by the HEPScore23 benchmark, either delivering the same capacity over the
whole 20 year time period, or a scenario with 15 percent year on year growth
to reflect an increase in computing required for scientific tasks.

We assume a load factor of 0.8 over the full lifetime of a compute node
compared to the load under a single benchmark run, the other time being
spent on software and hardware maintenance, batch system fill rates,
IO wait, etc. This is in line with what CERN reports for their batch system\cite{cernITsust}.

We use PUE data from existing facilities obtained from facilities managers
to explore the runtime carbon emissions for power and cooling in several
locations around the world, as well as a couple of hypothetical scenarios.

We simulate these facilities together with four strategies for hardware
replacement:

\begin{itemize}
    \item Replace after 3 years, or the shorter end of typical service contract
    \item Replace after 5 years, or the longer end of a typical service contract
    \item Replace after 10 years, or an estimate for average lifetime for ``run the hardware until it breaks''
    \item Don't replace hardware, but run it for the full 20 years (not likely with today's hardware construction)
\end{itemize}

\subsection{LCA model input data\label{subsec:Input-data}}

\subsubsection{Power generation emissions}

The power emissions have been taken from the 2023 average emissions for that
electricity zone from electricitymaps.com and we assume that this will
remain constant during the 20 year period simulated. 

\subsubsection{Cooling emissions}

In all but one centre, the cooling comes from compressors and has been
provided as PUE numbers. At the HPC2N centre, the cooling comes from district
cooling and has been provided as a direct carbon cost from the facility
annual sustainability report.

\subsubsection{Facility numbers -- real world\label{subsubsec:facilities}}

The real world facility numbers have been provided in direct communication
with the authors from the following centres:

\begin{itemize}
    \item ASGC: Academia Sinica Grid Computing Centre, Taipei, Taiwan
    \item BNL: Brookhaven National Lab, Long Island, USA
    \item HPC2N: High Performance Computing Centre North, Umeå, Sweden
    \item Vega: Vega EuroHPC, Maribor, Slovenia
\end{itemize}

For each facility we have the following information:

\begin{itemize}
    \item Yearly average electricity carbon intensity in 2023 from \cite{OWiD2024a}
    \item Electricity carbon intensity trend
    \item PUE
\end{itemize}

\subsubsection{Hypothetical scenario for heat reuse\label{subsubsec:heat-reuse}}

The `HPC2N heat reuse' scenario is constructed to explore what could be done with heat reuse, if we ignore costs and only focus on carbon emissions. Here we start with the `HPC2N' scenario, but instead of using district cooling we assume a big heat pump solution which increases PUE (and thus electricity use) but in turn reduces the use of district heating at the Umeå University campus during the winter (the campus buildings have a net heating need for about 6 months per year).

The carbon cost of district heating is 65 g/kWh and is mainly sourced from burning waste and wood. A heat pump with COP 2/3 is assumed, it consumes 1 kW of electricity and (together with the lukewarm returns) produces 2kW of cooling and 3 kW of heating increases the PUE from 1.03 to 1.33. The explicit average emissions per kWh cooled is then based on the kWh of district heating displaced $(65*2/3)/2$, or 48.75 g/kWh. In the model we have conservatively rounded this down to 40 g/kWh of negative emissions, based on it being a coarse hypothetical heat pump and not a system designed and studied in detail.

It is also possible that by redesigning the heating system in existing buildings for low temperature water heating, a more efficient heat reuse system could be designed, and it would be an interesting future work to explore this in more detail and include practical concerns like cost and reliability. Conversely, the HPC centre could employ hot water cooling \cite{ZIMMERMANN2012237} which would allow buildings to make use of the rest heat directly without the need for an additional heat pump.

\section{LCA simulations\label{sec:Calculations} }

In this section we explore various aspects of the cumulative emission calculations using our LCA model. We first calculate the emissions for a number of HPC centres in different locations and then more in detail at the effects of the embodied carbon of the hardware the server lifetime, the electricity generation carbon intensity and the facility expansion rate on the overall emissions. We present an analysis of the potential emission reductions from heat reuse.

\subsection{Cumulative emissions for various sites\label{subsec:cumulative-emissions-sites} }
\begin{figure*}[!h]
\includegraphics[width=0.9\textwidth]{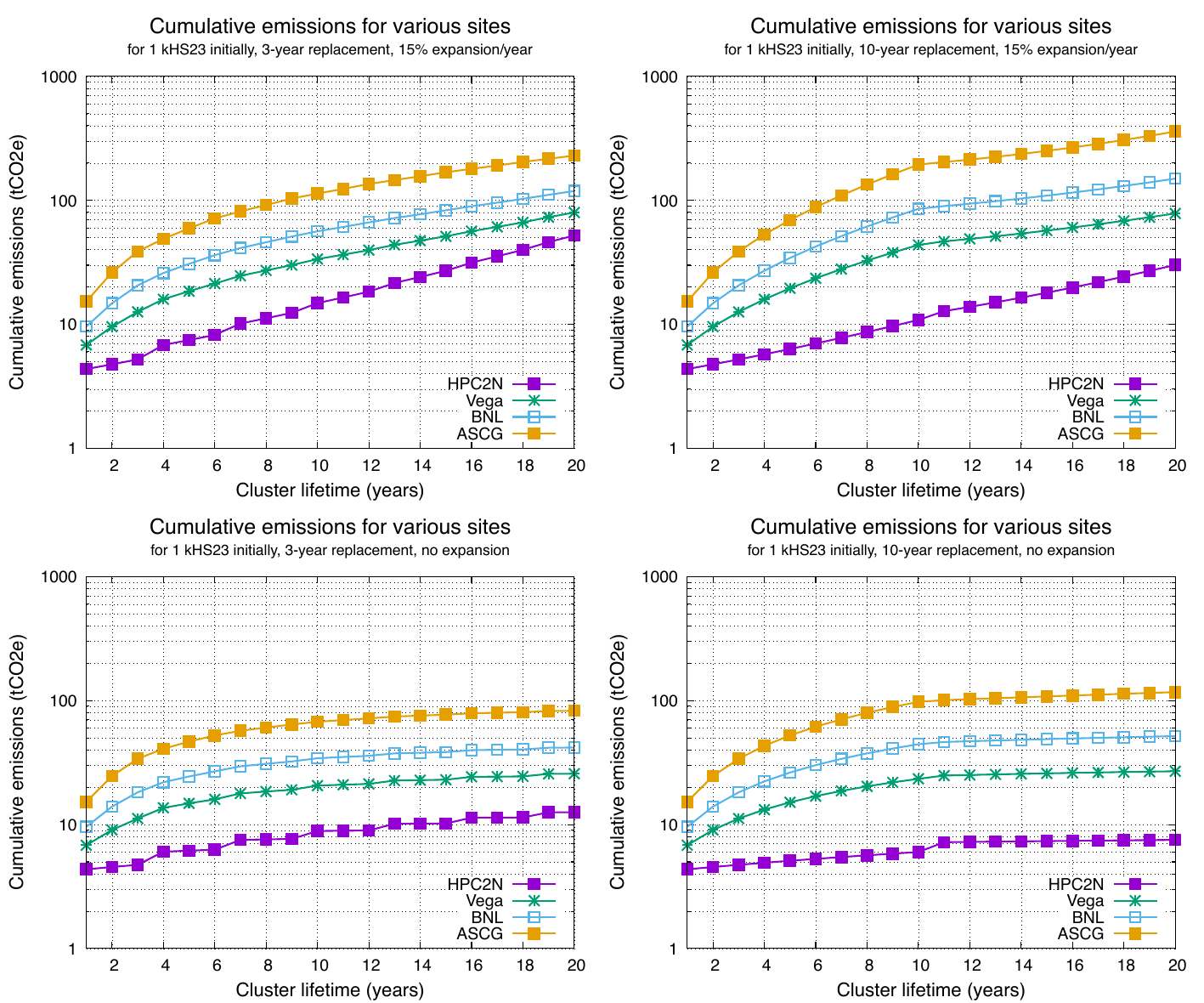}
\caption{Effect of location on cumulative emissions}
\label{fig:effect-sites}
\end{figure*}

Figure \ref{fig:effect-sites} show the emissions for the four facilities listed in \ref{subsubsec:facilities}. The site data are shown in Table \ref{tab:Site-data}. We assume that other factors such as kHS23 per node, server embodied carbon, server lifetime and facility expansion are the same for all sites. 

\begin{table}[h]
\begin{centering}
\begin{tabular}{|c|c|c|c|}
\hline 
Facility & CI (gCO$_{2}$/kWh) & $\Delta$CI/year & PUE\tabularnewline
\hline 
\hline 
HPC2N & 17 & 0.01 & 1.03\tabularnewline
\hline 
Vega & 231 & 0.05 & 1.13\tabularnewline
\hline 
BNL & 369 & 0.03 & 1.35\tabularnewline
\hline 
ASCG & 642 & 0.005 & 1.62\tabularnewline
\hline 
\end{tabular}
\par\end{centering}
\caption{Site data\label{tab:Site-data}}

\end{table}

Note that the figure uses a logarithmic scale for the Y-axis (cumulative emissions). What the figures show is that overall, the biggest contributor to growth in emissions is the cluster expansion: without expansion, energy efficiency gains and reductions in embodied carbon and in electricity carbon intensity result in the cumulative emissions flattening out; but even the modest 15\% growth results in a very considerable growth in overall emissions. 

for HPC2N, which is located in a region with low electricity carbon intensity (Sweden), the embodied carbon dominates and therefore longer replacement cycles reduce emissions. This is discussed in more detail in Sec. \ref{subsec:effect-ci}. 

The results also show the need for reducing electricity carbon intensity. The model includes the reduction trend for each country. It is clear that none of the three other countries (US, Taiwan and Slovenia) are reducing their low electricity carbon intensity fast enough to have an appreciable effect on the facility emissions.

\subsection{Effect of embodied carbon\label{subsec:effect-SSD} }

\subsubsection{Increasing SSD size}

\begin{figure*}[!h]
\centerline{\includegraphics[width=0.9\textwidth]{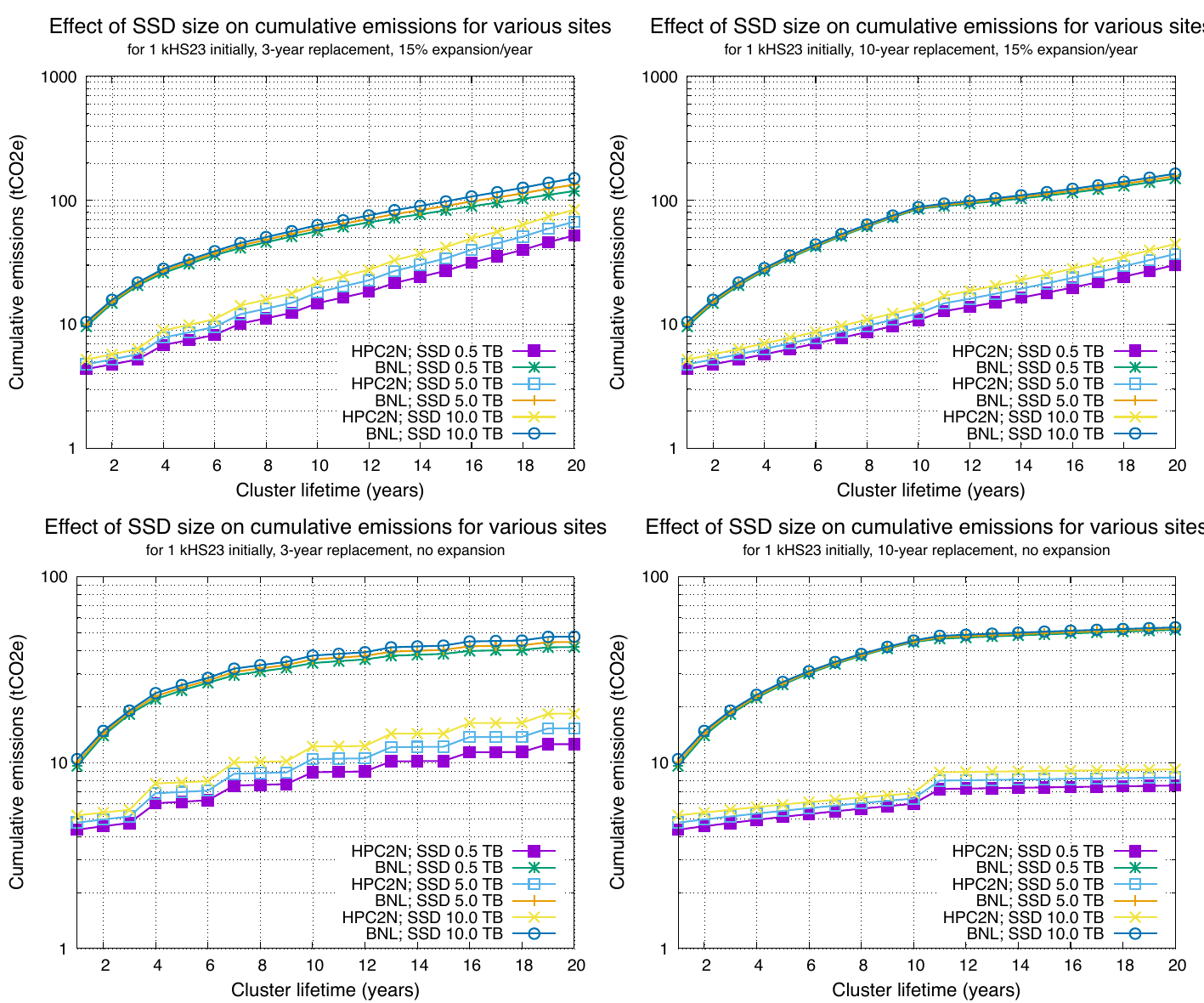}}
\caption{Effect of SSD size on cumulative emissions}
\label{fig:effect-SSD}
\end{figure*}

Throughout the paper we use the server configuration as described in Section \ref{subsec:Assumptions}, with an SSD of 0.5 TB. SSDs have a strong contribution to the embodied carbon \cite{10.1145/3630614.3630616}. To investigate the effect, we changed the SSD size to 5 and 10 TB for the three scenarios shown in Figure \ref{fig:effect-SSD}, for sites HPC2N and BNL. For a 0.5 TB SSD, the server embodied carbon is 1,200 kgCO$_2$e; for a 10 TB SSD, it is 2,042 kgCO$_2$e, so the SSD contributes 88.6 kgCO$_2$e/TB. 

\subsubsection{Embodied carbon trend}\label{Sec:embodied-carbon-trend}

According to \cite{10.1145/3630614.3630616}, the increase in emissions per unit area from the 28 nm CMOS node (2010) to the 3 nm node (2022) is $3\times$, or about 10\% year-on-year increase. The compute performance per unit area improves as feature size shrinks, so that in that same period, it improved $(28/3)^2=87\times$.  
The density (bits/unit area) increases with every generation. \cite{electronics10243156} shows that density increased $100\times$ between 2008 and 2020. These trends have been slowing down slightly as we reach the end of feature size scaling on CMOS. But if we consider the storage density and compute performance for a constant amount of embodied carbon, then in the past 12 years we have seen a 30\% reduction year-on-year. 

In practice, when servers are upgraded, SSDs are usually replaced by larger ones, and RAM size and CPU performance increase as well. As a result, the embodied emissions will not decrease much, if at all. As the actual die sizes depend on commercial factors and are therefore not following a clear trend, we assume for simplicity that the increase in density compensates the increase capacity and emissions per unit area so that there is not much change in the effect of the embodied carbon over time. We make similar assumptions for the memory and compute chips. If the CI is high, the SSD embodied carbon contribution is small, so a small or large SSD does not make much of a difference; in low-CI regions, the embodied carbon makes up most of the total cumulative emissions, and the difference between a small and a large SSD is up to 60\% on the total figure.

\begin{figure*}[!h]
\centerline{\includegraphics[width=0.9\textwidth]{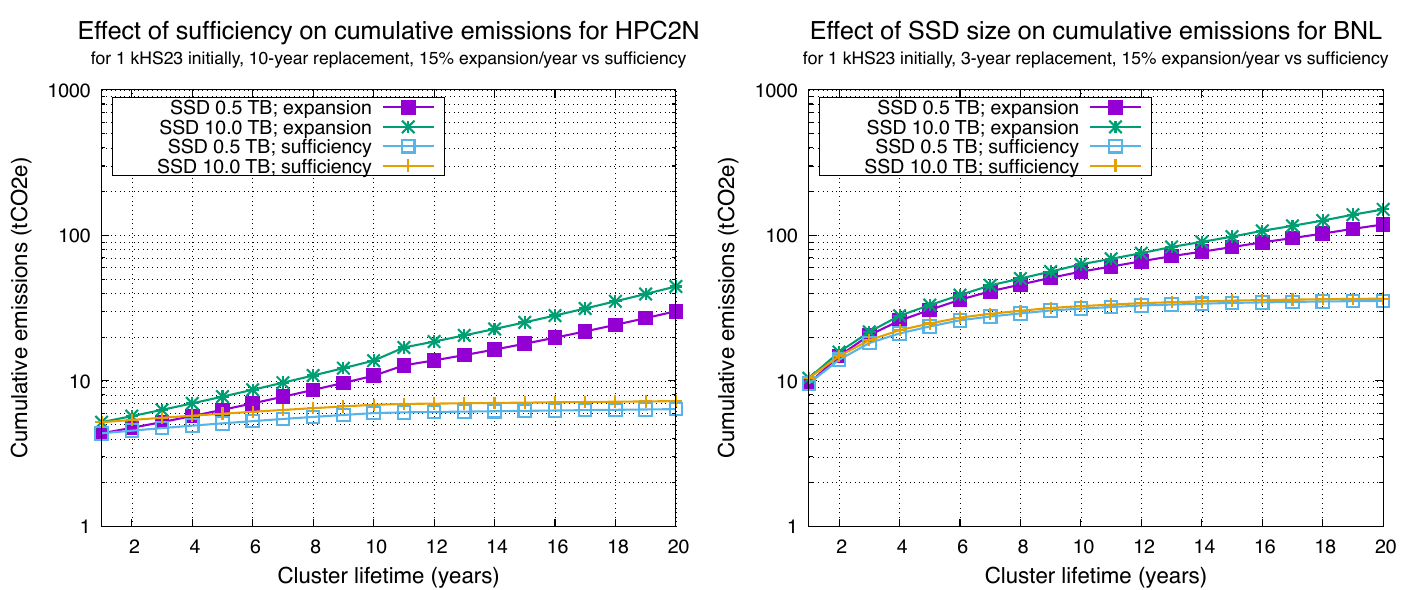}}
\caption{Effect of sufficiency assumption combined with embodied emissions trend on cumulative emissions}
\label{fig:effect-ec-trend}
\end{figure*}

However, if an HPC centre would opt for \emph{sufficiency} \cite{zora110766}, i.e. not grow its compute capacity or storage on consecutive replacements, the reduction in embodied carbon would be very significant, as shown in Fig. \ref{fig:effect-ec-trend}. We show HPC2N with 10-year replacement and BNL with 3-years replacement, for 0.5 TB and 10 TB SSD size. The "expansion" scenario assumes cluster expansion in terms of kHS23 as well as scaling up of SSD size and RAM size. The "sufficiency" scenario assumes the cluster capacity remains constant in terms of kHS23, memory and storage.

\subsection{Effect of server lifetime\label{subsec:effect-lifetime} }

\begin{figure}[h!]
\centerline{\includegraphics[width=0.95\columnwidth]{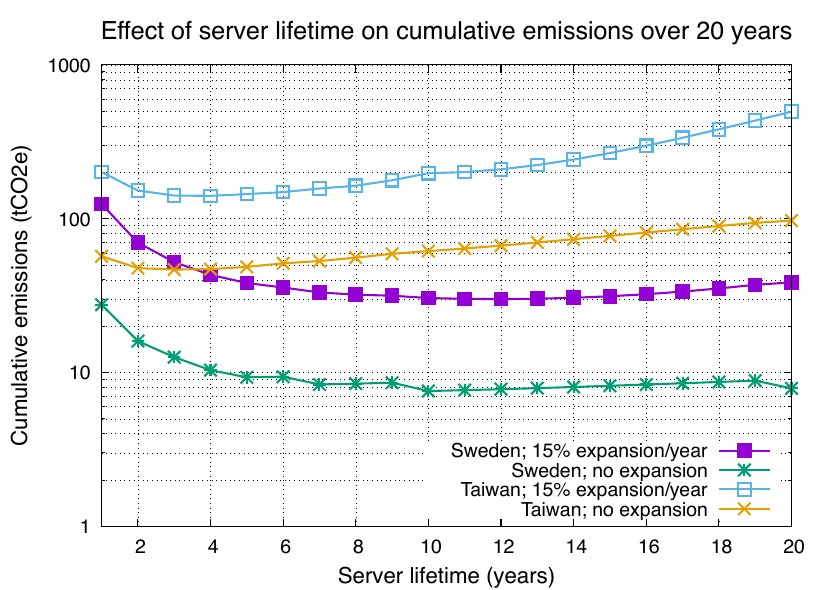}}
\caption{Effect of server lifetime on cumulative emissions}
\label{fig:effect-lifetime}
\end{figure}

To gain a better understanding of the effect of server lifetimes, we looked at the emissions as a function of the lifetime for two countries, one with very low CI (Sweden) and one with a very high CI (Taiwan), with and without 15\% cluster expansion per year. The results are shown in Figure \ref{fig:effect-lifetime}.

For the high-CI case, faster replacement is better as energy efficiency reduces emissions and embodied carbon is a small contribution. 

For the low-CI case, the interesting observation is that there is an optimal lifetime at which the combined contributions of embodied carbon and emissions from use are minimised, and that this lifetime is around ten years.

\subsection{Effect of electricity carbon intensity\label{subsec:effect-ci} }

\begin{figure}[h!]
\centerline{\includegraphics[width=0.95\columnwidth]{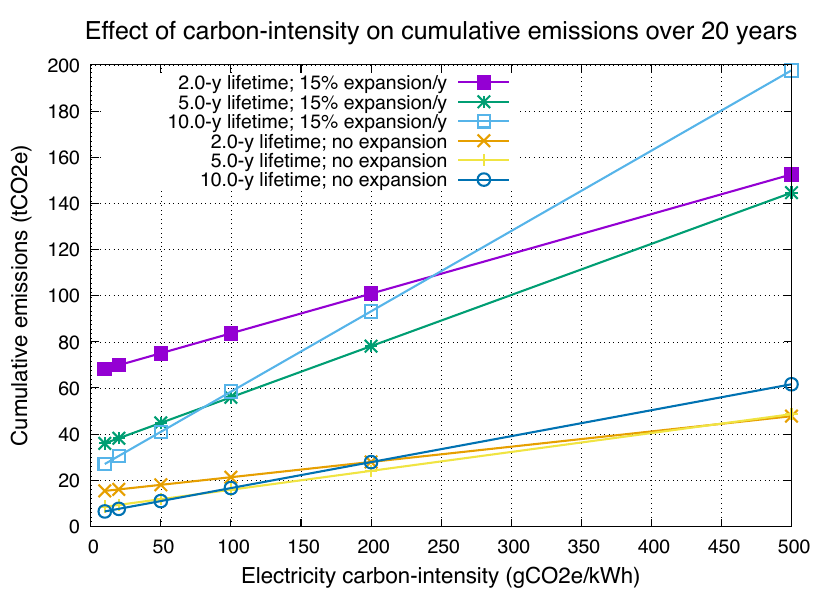}}
\caption{Effect of electricity carbon intensity on cumulative emissions}
\label{fig:effect-ci}
\end{figure}

To study the effect of electricity carbon intensity, we vary the CI and show cumulative emissions for  2, 5 and 10 year replacement, with and without expansion. The results are shown in Figure \ref{fig:effect-ci}. We see that the emissions increase linear with the CI, and that with increasing CI there is a crossover point for the various replacement cycles. This is because for low CI, embodied carbon dominates, so that longer replacement cycles are better, but for high CI, it is the opposite. The difference between 2-year and 5-year cycles in that case is small because the energy efficiency increases only slowly.

\begin{figure}[h!]
\centerline{\includegraphics[width=0.95\columnwidth]{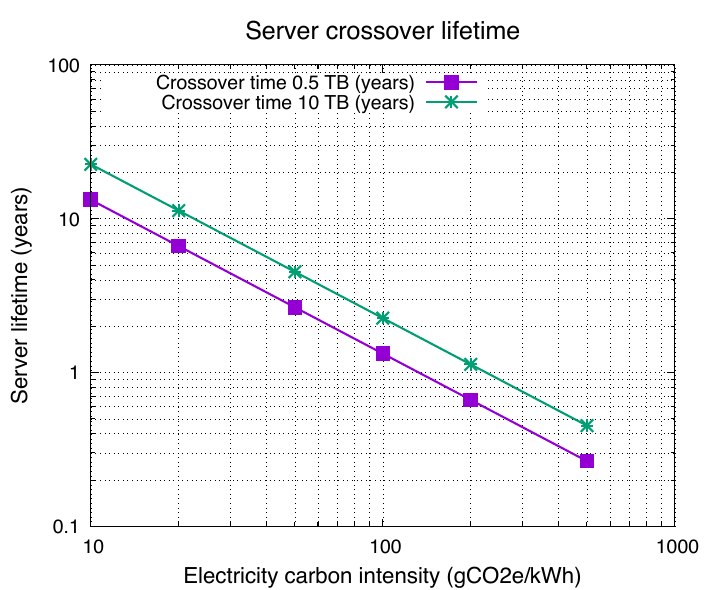}}
\caption{Effect of electricity carbon intensity on cross-over from embodied emissions to emissions from use}
\label{fig:ci-crossover}
\end{figure}

We can also compare the relative importance of embodied carbon and emissions from use, as shown in Figure \ref{fig:ci-crossover}. The figure shows the time at which the emissions from use of the exemplar server used in this work equal the embodied carbon as a function of the electricity carbon intensity of the region where the facility is located. As the emissions from use grow linearly with the CI, and the embodied carbon does not grow (no replacement and no expansion), the crossover lifetime is also a linear function of the CI, as shown in Eq. \ref{eqCrossover}.
\begin{equation}
\mathtt{crossover\_lifetime}(CI) =  \frac{node_{ec}}{CI.E_{node}} \label{eqCrossover}
\end{equation}

We use two different SSD sizes, 0.5 TB and 10 TB. For regions with very low CI, the server lifetime for break even is already quite long with the smaller SSD (close to 15 years); with a 10 TB SSD, the server would need to run for more than 20 years to reach the crossover point. For regions with high CI, the relative contribution of the embodied carbon is small so emissions from use soon dominate. Even for a region with relatively low CI of around 200 gCO$_2$e/kWh such as the UK, we see that the crossover lifetime is shorter than the typical current useful life, and the addition embodied carbon from the larger SSD does not change this much. For a CI of 40 gCO$_2$e/kWh (e.g. France), the crossover lifetime for the server with the large SSD is around 5 years, which is of the same order as current useful life, and for a CI of 20 gCO$_2$e/kWh (e.g. Sweden), it would be 10 years. 

\subsection{Effect of facility expansion rate\label{subsec:effect-expansion} }

\begin{figure}[h!]
\centerline{\includegraphics[width=0.95\columnwidth]{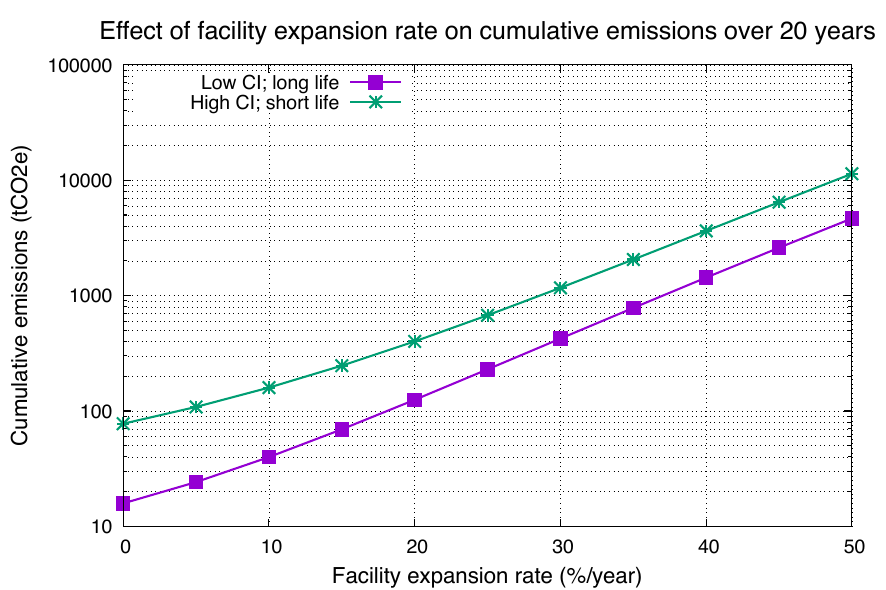}}
\caption{Effect of facility expansion rate on cumulative emissions}
\label{fig:effect-expansion}
\end{figure}

To illustrate the effect of the facility expansion rate we vary the expansion for two cases: low CI, 10-year replacement cycle (`long life') and high CI,  2-year replacement cycle (`short life'). The results are shown in Figure \ref{fig:effect-expansion}. Note that the cumulative emissions are shown on a logarithmic scale, because a fixed yearly expansion rate results in exponential expansion as per Equation \ref{eq2} and therefore exponential growth in emissions.

\begin{figure*}[t!]
\includegraphics[width=0.9\textwidth]{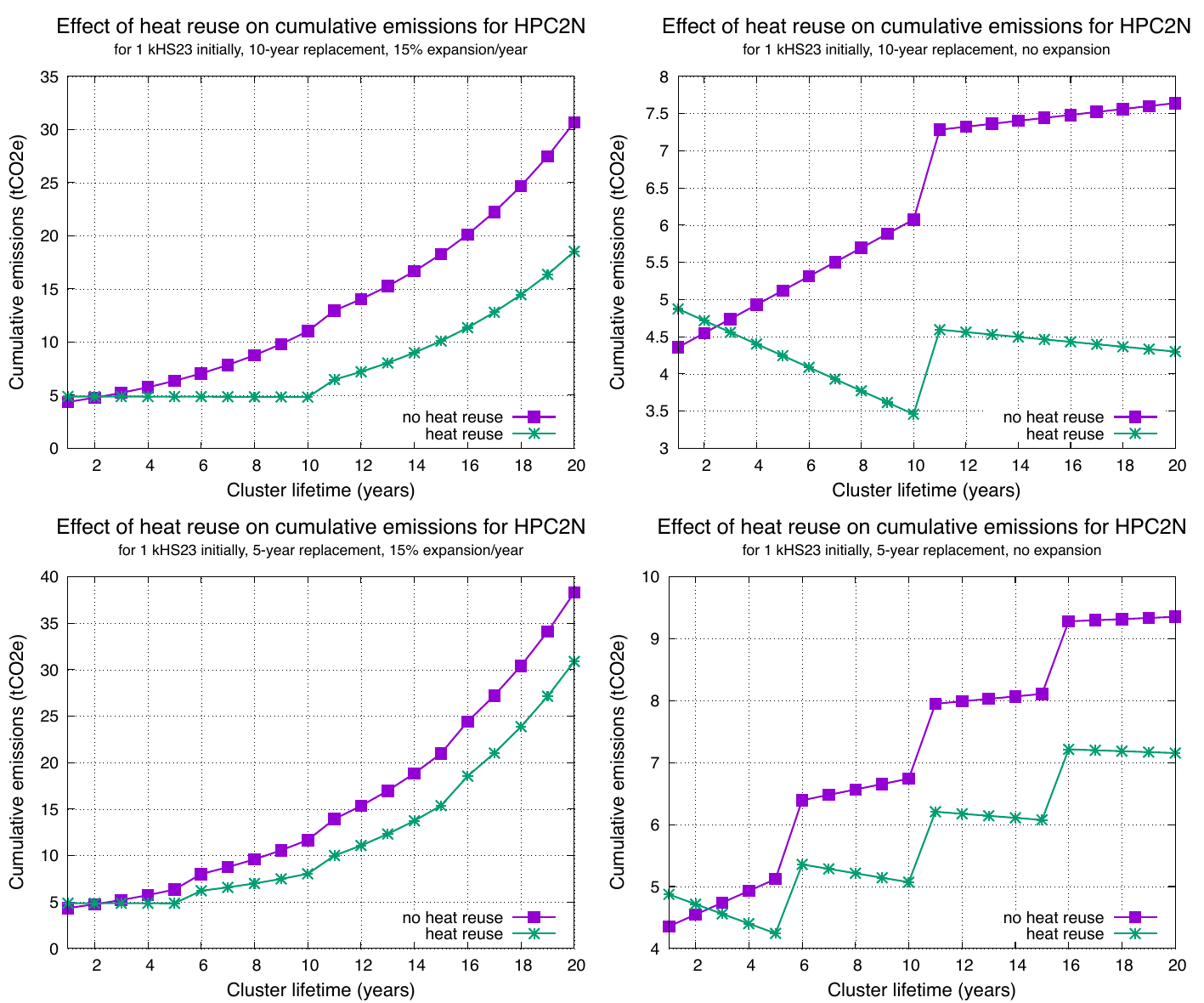}
\caption{Effect of heat reuse on cumulative emissions for HPC2N}
\label{fig:effect-heat-reuse}
\end{figure*}

\subsection{Effect of heat reuse\label{subsec:effect-heat-reuse} }

For the heat reuse scenario discussed under Section \ref{subsubsec:heat-reuse}, we consider replacement cycles of 5 and 10 years, with and without expansion, for the HPC2N site with and without the hypothetical heat reuse. The results are shown in Figure \ref{fig:effect-heat-reuse}. The most striking result is the case of a 10-year replacement cycle without cluster expansion. In this case, the heat reuse scenario results in close to zero cumulative emissions over 20 years. In all cases, heat reuse results in considerable emission reductions. 

The figures also illustrate quite clearly that even in a low-emission region with heat reuse, the cluster expansion will result in very considerable cumulative emissions over its lifetime.

\begin{figure*}[!t]
\centerline{\includegraphics[width=0.9\textwidth]{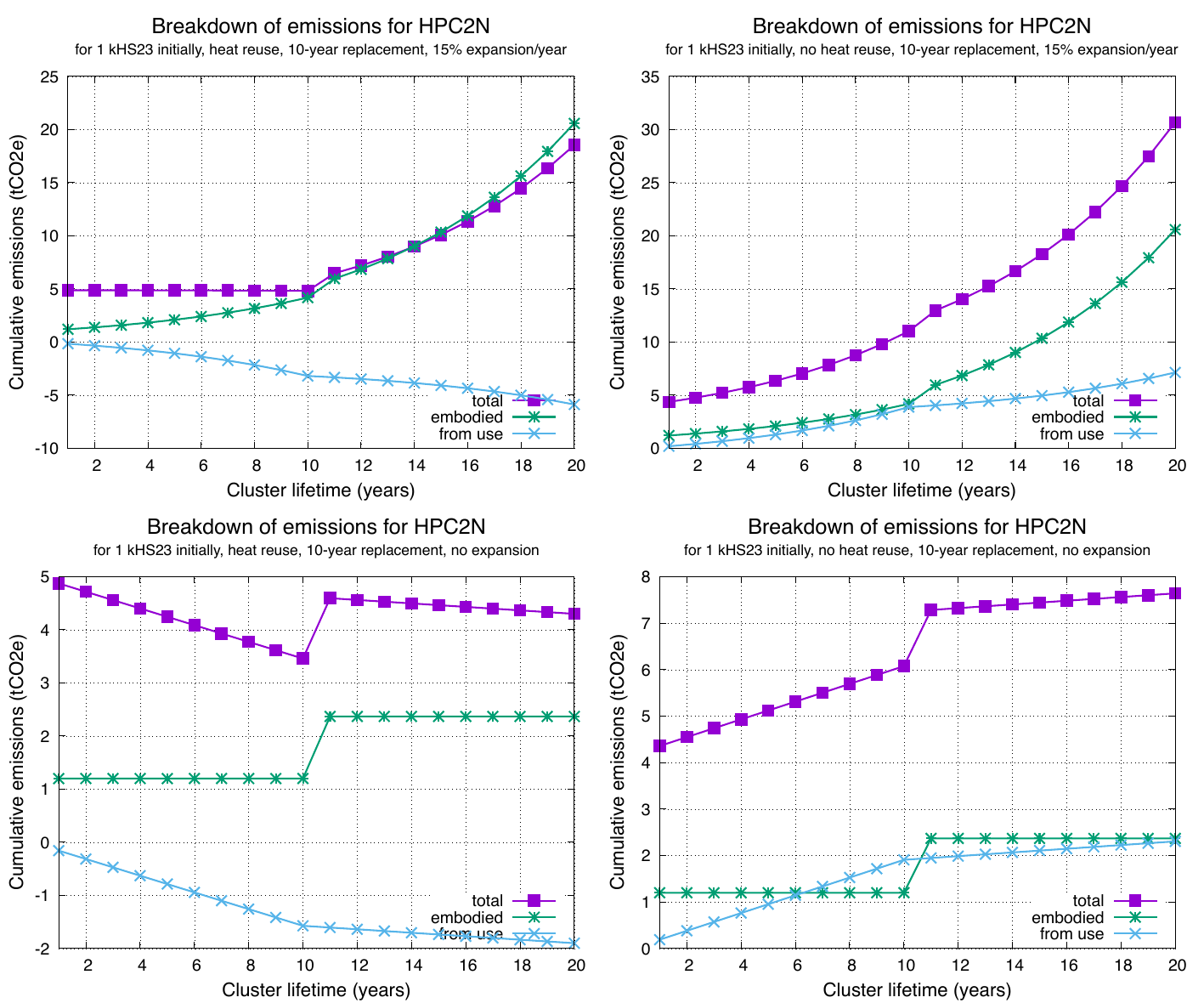}}
\caption{Breakdown of cumulative emissions for HPC2N with and without heat reuse}
\label{fig:effect-heat-reuse-expl}
\end{figure*}

The jump in cumulative emissions at the replacement year may seem high at first sight. To better explain this behaviour, Figure \ref{fig:effect-heat-reuse-expl} shows the breakdown into embodied emissions and emissions from use with and without heat reuse, for the case of replacement after 10 years and 15\% expansion. The behaviour becomes more clear if we focus on the embodied carbon emissions: without expansion, the cumulative embodied emissions increase on replacement of the server, after 10 years. The emissions from use are negative because of the heat reuse, resulting in the jagged shape of the total cumulative emissions curve.
With 15\% expansion the embodied carbon increases gradually over 9 years. In the 10th year the initially installed nodes of this cluster are replaced by new ones that have higher performance per Watt, so the embodied emissions jump with that amount. Then they grow through yearly expansion as well as the yearly replacement of the nodes installed 10 years before. So the  embodied emissions curve is a gradual upward curve with a slight kink after 10 years.

\subsection{Discussion\label{sec:Discussion}}

The above calculations serve to exercise the capabilities of our LCA model as well as to illustrate the key drivers for overall emissions. There are a few general observations. Cluster expansion to account for projected growth in compute needs has the highest impact on final emissions: with 50\% expansion, the cumulative emissions over 20 years are at least two orders of magnitude higher than without expansion. Shorter replacement cycles result in considerably higher embodied carbon. A considerable factor in this is the size of the SSD: a 10 TB SSD results in up to 60\% higher cumulative embodied emissions than a 0.5 TB SSD. In high-CI regions there is an optimum lifetime; in low-CI regions, embodied carbon quickly becomes the dominant contribution to overall emissions. To minimise emissions, server lifetimes will need to be much longer than current (up to 20 years, similar to the lifetime of the data centre itself). It also follows that, to minimise emissions for a given compute budget, low-CI facilities could consider using end-of-life hardware from facilities in carbon-intensive regions. The heat reuse scenario shows the potential benefits of using the facility waste heat for district heating or other heating purposes, e.g. swimming pools or greenhouses.

\section{Validation}

To validate our model, we tried to reproduce the values from the other
LCA analysis papers. We could not reproduce the values from paper
\cite{samaye2025LCA}, as the information provided on the assumptions is incomplete.
Paper \cite{whitehead2015LCA} expresses all values in Eco-indicator points, and there
is not sufficient information in the paper to separate the emissions
from the other factors contributing to the points.

Table \ref{table:Ma} shows the comparison between our model and that of \cite{Ma2024LCA}. In that paper, Fig. 6. shows the emissions for three sites: Shenzen (China), Guizhou (China) and North Virginia (USA). The values in the figure are taken from Table 2, which shows the yearly emissions.
The papers gives 62.782 TWh/year for the ``Total energy consumption''
values in the table. However, the discussion mentions that the data
centre is assumed to host 21,000 Dell PowerEdge R710 2U rack servers.

\begin{table*}[!h]
\begin{centering}
\begin{tabular}{|l|c|c|c|}
\hline 
LCA calculation & Shenzen & Guizhou & North Virginia\tabularnewline
\hline 
\hline 
Yearly emissions from \cite{Ma2024LCA}  & 31.277 & 28.539 & 16.833\tabularnewline
\hline 
Reproduced with our model, year 1 & 31.3 & 28.6 & 16.9\tabularnewline
\hline 
\hline 
Our model, specs from \cite{Ma2024LCA}, with replacement, year 1 & 21.5 & 19.8 & 12.1\tabularnewline
\hline 
Our model, specs from \cite{Ma2024LCA}, with replacement, 20-year average & 8.2 & 7.4 & 4.3\tabularnewline
\hline 
\hline 
Our model, our specs, with replacement, year 1 & 22.7 & 20.9 & 13.3\tabularnewline
\hline 
Our model, our specs, with replacement, 20-year average & 8.4 & 7.7 & 4.5\tabularnewline
\hline 
\hline
Our model, our specs, matched expansion, year 1 & 22.7 & 20.9 & 13.3\tabularnewline
\hline 
Our model, our specs, matched expansion, 20-year average & 30.0 & 27.8 & 18.0\tabularnewline
\hline 
\end{tabular}
\par\end{centering}
\caption{Comparison of our model with that of \cite{Ma2024LCA}. All values are in million kgCO$_{2}$/year}\label{table:Ma}
\end{table*}

\begin{figure*}[h!]
\centerline{\includegraphics[width=0.7\textwidth]{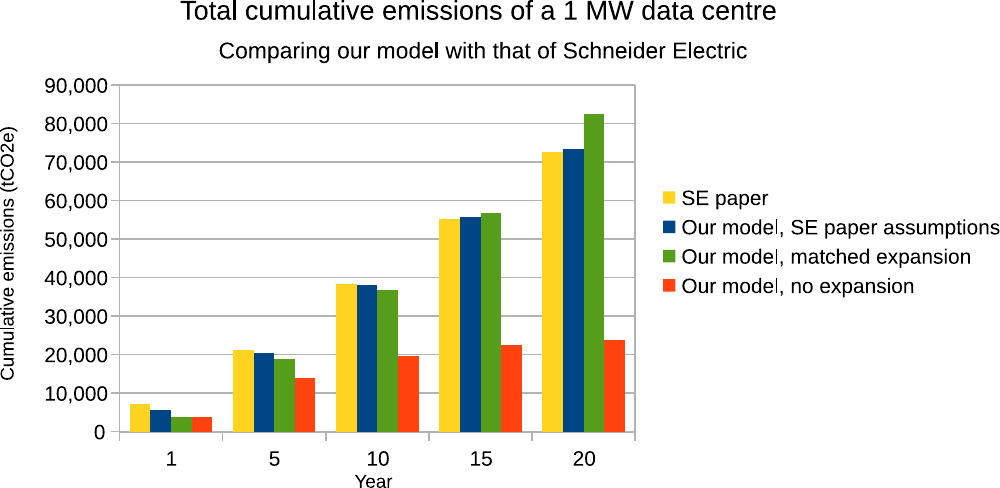}}
\caption{Comparison of our model with that of \cite{schneider2023}}
\label{fig:ours-vs-SE}
\end{figure*}

These servers have a power consumption of between 200 W and 300 W.
From the values in the paper, the calculated power consumption would
be 62.782/(365$\times$24) TW or 7.16 GW. Per server this would amount to
341.3 kW. We assume that this is an SI prefix error (T instead of
G), and that the actual data centre draws 7.16 MW and therefore has
an energy consumption of 63.782 GWh/y. If that is correct, and taking
into account the PUE, the server power consumption would be 216 W.

With those assumptions, our model reproduces the yearly emissions
from the paper: resp. 31.3, 28.6, 16.9 million kgCO$_{2}$/year. This
paper ignores efficiency gains and embodied carbon increases due to
server replacement. The total value of 159 kgCO$_{2}$e/y embodied
carbon for the server is on the low side. The paper also either implicitly
assumes a server load factor of 1, even though there is a load factor
$\eta^{p}$ in the equations. There is no factor for the idle power
consumption. 

To improve on this, we assume the same data centre hardware specs but
a load factor of 50\% and an idle power consumption of 30\%. If we
calculate the cumulative emissions over 20 years and consider those
per year, we get emissions of only 8.2, 7.4, 4.3 million kgCO$_{2}$/year; purely for the first year it would be 21.5, 19.8, 12.1. 

We can further refine the estimate by applying the assumptions used in our own work for the server specification and the data centre infrastructure. We scale the number of nodes
so that $n_{nodes}.P_{node}.PUE=7.16MW$. This results in 8.4, 7.7, 4.5 million
kgCO$_{2}$/year; purely for the first year it would be 22.7, 20.9,
13.3. The differences between our specs and those from \cite{Ma2024LCA} result in small differences in
emissions because all three sites have a comparatively high electricity
carbon intensity. The main difference arises from the calculation of the cumulative emissions with replacement, which results in much lower yearly emissions when averaged over the data centre lifetime.

Instead of like-for-like replacement, which reduces the server power consumption, we can also assume matched expansion, which increases the compute power and maintains the power consumption. These values are shown in the last two rows of the table.

Figure \ref{fig:ours-vs-SE} shows the comparison between our model and that of \cite{schneider2023}.
In that paper, Fig. 3(a) shows the cumulative total carbon footprint
profile over time. We reproduced the total values in this graph based
on the ``Main assumptions'' in the paper. The paper assumes that
the total power consumption of the IT equipment is 500 kW (their Table
A1) and does not consider server load or idle power consumption. Consequently,
we set the load to 100\%. Furthermore, although the model includes
server replacement every 4 years, it does not account for the efficiency
gains of the servers, nor for the decrease of the embodied carbon
or of the electricity carbon intensity. The paper also assumes a very
high electricity carbon intensity of 511 gCO$_{2}$e/kWh. This value
does not have a source attribution. Finally, the paper assumes a rather
ad-hoc embodied carbon contribution 5 kgCO$_{2}$e per Watt of installed
capacity, again without attribution. 

We first reproduced the values
from the paper using their assumptions to show its validity, and then
improved on their estimate using our assumptions. We used the same
CI value as for the BNL data centre, 369 gCO$_{2}$e/kWh, which is
the US average value for 2023 reported by \cite{OWiD2024a}. We kept the PUE
value of 1.34 as assumed in the paper. The other assumptions are the
same as in \ref{subsec:Assumptions}. The resulting emissions are smaller and on
a clearly decreasing trend. These key differences are caused by the
energy efficiency gains of every next generation of servers and the
lower electricity carbon intensity. 

As in the previous scenario, instead of like-for-like replacements, which reduces the server power consumption, we can also assume matched expansion, which increases the compute power and maintains the power consumption. In this scenario, although the energy consumption does not increase, the embodied carbon increases compared to the assumptions from \cite{schneider2023}.   

In summary, our model is not only able to accurately reproduce the values from other models but allows to improve considerably on existing LCA calculations for data centres. 

\section{Conclusions}

The question if an HPC site is better off replacing old compute hardware with more efficient newer hardware, or keep running the old hardware for as long as practically possible depends on the local power carbon intensity. Using our total emissions simulation tool, site administrators can make better informed decisions with regards to lifecycle carbon impact.

With advancing decarbonisation of the electricity grid, most HPC sites will eventually be in the regime where emissions are dominated by hardware replacement cycles and expansion. Reducing the carbon emissions from computer hardware manufacturing should be a focus area as much as reducing emissions from electricity generation. Ever-growing expansion will have ever-growing emissions, even in places with low carbon intensity electricity, so it becomes important to question if expansion is necessary for the science and justifiable in terms of emissions.

\bibliography{paper-LCA-for-HPC}


\begin{thebibliography}{23}
\ifx \bisbn   \undefined \def \bisbn  #1{ISBN #1}\fi
\ifx \binits  \undefined \def \binits#1{#1}\fi
\ifx \bauthor  \undefined \def \bauthor#1{#1}\fi
\ifx \batitle  \undefined \def \batitle#1{#1}\fi
\ifx \bjtitle  \undefined \def \bjtitle#1{#1}\fi
\ifx \bvolume  \undefined \def \bvolume#1{\textbf{#1}}\fi
\ifx \byear  \undefined \def \byear#1{#1}\fi
\ifx \bissue  \undefined \def \bissue#1{#1}\fi
\ifx \bfpage  \undefined \def \bfpage#1{#1}\fi
\ifx \blpage  \undefined \def \blpage #1{#1}\fi
\ifx \burl  \undefined \def \burl#1{\textsf{#1}}\fi
\ifx \doiurl  \undefined \def \doiurl#1{\url{https://doi.org/#1}}\fi
\ifx \betal  \undefined \def \betal{\textit{et al.}}\fi
\ifx \binstitute  \undefined \def \binstitute#1{#1}\fi
\ifx \binstitutionaled  \undefined \def \binstitutionaled#1{#1}\fi
\ifx \bctitle  \undefined \def \bctitle#1{#1}\fi
\ifx \beditor  \undefined \def \beditor#1{#1}\fi
\ifx \bpublisher  \undefined \def \bpublisher#1{#1}\fi
\ifx \bbtitle  \undefined \def \bbtitle#1{#1}\fi
\ifx \bedition  \undefined \def \bedition#1{#1}\fi
\ifx \bseriesno  \undefined \def \bseriesno#1{#1}\fi
\ifx \blocation  \undefined \def \blocation#1{#1}\fi
\ifx \bsertitle  \undefined \def \bsertitle#1{#1}\fi
\ifx \bsnm \undefined \def \bsnm#1{#1}\fi
\ifx \bsuffix \undefined \def \bsuffix#1{#1}\fi
\ifx \bparticle \undefined \def \bparticle#1{#1}\fi
\ifx \barticle \undefined \def \barticle#1{#1}\fi
\bibcommenthead
\ifx \bconfdate \undefined \def \bconfdate #1{#1}\fi
\ifx \botherref \undefined \def \botherref #1{#1}\fi
\ifx \url \undefined \def \url#1{\textsf{#1}}\fi
\ifx \bchapter \undefined \def \bchapter#1{#1}\fi
\ifx \bbook \undefined \def \bbook#1{#1}\fi
\ifx \bcomment \undefined \def \bcomment#1{#1}\fi
\ifx \oauthor \undefined \def \oauthor#1{#1}\fi
\ifx \citeauthoryear \undefined \def \citeauthoryear#1{#1}\fi
\ifx \endbibitem  \undefined \def \endbibitem {}\fi
\ifx \bconflocation  \undefined \def \bconflocation#1{#1}\fi
\ifx \arxivurl  \undefined \def \arxivurl#1{\textsf{#1}}\fi
\csname PreBibitemsHook\endcsname

%%% 1
\bibitem[\protect\citeauthoryear{Vanderbauwhede and
  Stubben}{2023}]{wim_vanderbauwhede_2023_7709483}
\begin{botherref}
\oauthor{\bsnm{Vanderbauwhede}, \binits{W.}},
\oauthor{\bsnm{Stubben}, \binits{O.}}:
Impact of the development and deployment of software on energy consumption:
  findings and recommendations.
Zenodo
(2023).
\doiurl{10.5281/zenodo.7709483}
\end{botherref}
\endbibitem

%%% 2
\bibitem[\protect\citeauthoryear{Lord et~al.}{2025}]{10.1145/3706598.3713919}
\begin{bchapter}
\bauthor{\bsnm{Lord}, \binits{C.}},
\bauthor{\bsnm{Friday}, \binits{A.}},
\bauthor{\bsnm{Jackson}, \binits{A.}},
\bauthor{\bsnm{Bird}, \binits{C.}},
\bauthor{\bsnm{Preist}, \binits{C.}},
\bauthor{\bsnm{Lambert}, \binits{S.}},
\bauthor{\bsnm{Kayumbi}, \binits{G.}},
\bauthor{\bsnm{Widdicks}, \binits{K.}}:
\bctitle{The world is not enough: Growing waste in hpc-enabled academic
  practice}.
In: \bbtitle{Proceedings of the 2025 CHI Conference on Human Factors in
  Computing Systems}.
\bsertitle{CHI '25}.
\bpublisher{Association for Computing Machinery},
\blocation{New York, NY, USA}
(\byear{2025}).
\doiurl{10.1145/3706598.3713919}
\end{bchapter}
\endbibitem

%%% 3
\bibitem[\protect\citeauthoryear{Ma and Zhou}{2024}]{Ma2024LCA}
\begin{barticle}
\bauthor{\bsnm{Ma}, \binits{K.}},
\bauthor{\bsnm{Zhou}, \binits{Y.}}:
\batitle{A comprehensive quantitative lifecycle cost and environmental impact
  analysis model for computing infrastructure}.
\bjtitle{MethodsX}
\bvolume{13},
\bfpage{103009}
(\byear{2024})
\doiurl{10.1016/j.mex.2024.103009}
\end{barticle}
\endbibitem

%%% 4
\bibitem[\protect\citeauthoryear{Whitehead et~al.}{2015}]{whitehead2015LCA}
\begin{barticle}
\bauthor{\bsnm{Whitehead}, \binits{B.}},
\bauthor{\bsnm{Andrews}, \binits{D.}},
\bauthor{\bsnm{Shah}, \binits{A.}}:
\batitle{The life cycle assessment of a uk data centre}.
\bjtitle{The International Journal of Life Cycle Assessment}
\bvolume{20},
\bfpage{332}--\blpage{349}
(\byear{2015})
\end{barticle}
\endbibitem

%%% 5
\bibitem[\protect\citeauthoryear{Samaye et~al.}{2025}]{samaye2025LCA}
\begin{botherref}
\oauthor{\bsnm{Samaye}, \binits{I.}},
\oauthor{\bsnm{Ouffou\'{e}}, \binits{G.}},
\oauthor{\bsnm{Gamati\'{e}}, \binits{A.}}:
Life cycle assessment of edge data centers: Case study in presence of renewable
  energy and refurbished servers.
ACM J. Comput. Sustain. Soc.
\textbf{3}(2)
(2025)
\doiurl{10.1145/3724127}
\end{botherref}
\endbibitem

%%% 6
\bibitem[\protect\citeauthoryear{Lin et~al.}{2023}]{schneider2023}
\begin{botherref}
\oauthor{\bsnm{Lin}, \binits{P.}},
\oauthor{\bsnm{Bunger}, \binits{R.}},
\oauthor{\bsnm{Avelar}, \binits{V.}}:
{Quantifying Data Center Scope 3 GHG Emissions to Prioritize Reduction Efforts}
(2023).
\url{https://www.se.com/ww/en/download/document/SPD_WP99_EN/}
Accessed 2025-06-04
\end{botherref}
\endbibitem

%%% 7
\bibitem[\protect\citeauthoryear{Vanderbauwhede and
  Wadenstein}{2025}]{hpc_lca_code_wv2025}
\begin{botherref}
\oauthor{\bsnm{Vanderbauwhede}, \binits{W.}},
\oauthor{\bsnm{Wadenstein}, \binits{M.}}:
{LCA model for servers in a data centre}
(2025).
\url{https://codeberg.org/wimvanderbauwhede/low-carbon-computing/src/branch/master/LCA-model-equations}
Accessed 2025-06-04
\end{botherref}
\endbibitem

%%% 8
\bibitem[\protect\citeauthoryear{Szczepanek et~al.}{2024}]{szczepanek2024hep}
\begin{barticle}
\bauthor{\bsnm{Szczepanek}, \binits{N.}},
\bauthor{\bsnm{Britton}, \binits{D.}},
\bauthor{\bsnm{Di~Girolamo}, \binits{A.}},
\bauthor{\bsnm{Ketele}, \binits{E.}},
\bauthor{\bsnm{Glushkov}, \binits{I.}},
\bauthor{\bsnm{Giordano}, \binits{D.}},
\bauthor{\bsnm{Ondris}, \binits{L.}},
\bauthor{\bsnm{Simili}, \binits{E.}},
\bauthor{\bsnm{Borge}, \binits{G.M.}}:
\batitle{Hep benchmark suite: Enhancing efficiency and sustainability in
  worldwide lhc computing infrastructures}.
\bjtitle{arXiv preprint arXiv:2408.12445}
(\byear{2024})
\doiurl{10.48550/arXiv.2408.12445}
\end{barticle}
\endbibitem

%%% 9
\bibitem[\protect\citeauthoryear{Koomey et~al.}{2011}]{koomey2011web}
\begin{barticle}
\bauthor{\bsnm{Koomey}, \binits{J.G.}},
\bauthor{\bsnm{Berard}, \binits{S.}},
\bauthor{\bsnm{Sanchez}, \binits{M.}},
\bauthor{\bsnm{Wong}, \binits{H.}}:
\batitle{Web extra appendix: implications of historical trends in the
  electrical efficiency of computing}.
\bjtitle{IEEE Annals of the History of Computing}
\bvolume{33}(\bissue{3}),
\bfpage{1}--\blpage{30}
(\byear{2011})
\end{barticle}
\endbibitem

%%% 10
\bibitem[\protect\citeauthoryear{Masanet
  et~al.}{2020}]{doi:10.1126/science.aba3758}
\begin{barticle}
\bauthor{\bsnm{Masanet}, \binits{E.}},
\bauthor{\bsnm{Shehabi}, \binits{A.}},
\bauthor{\bsnm{Lei}, \binits{N.}},
\bauthor{\bsnm{Smith}, \binits{S.}},
\bauthor{\bsnm{Koomey}, \binits{J.}}:
\batitle{Recalibrating global data center energy-use estimates}.
\bjtitle{Science}
\bvolume{367}(\bissue{6481}),
\bfpage{984}--\blpage{986}
(\byear{2020})
\doiurl{10.1126/science.aba3758}
\end{barticle}
\endbibitem

%%% 11
\bibitem[\protect\citeauthoryear{Ember}{2024}]{OWiD2024a}
\begin{botherref}
\oauthor{\bsnm{Ember}}:
{Carbon intensity of electricity generation -- Ember and Energy Institute}
(2024).
\url{https://ourworldindata.org/grapher/carbon-intensity-electricity}
Accessed 2025-01-20
\end{botherref}
\endbibitem

%%% 12
\bibitem[\protect\citeauthoryear{Lorenzini}{2021}]{lorenzini2021digital}
\begin{botherref}
\oauthor{\bsnm{Lorenzini}, \binits{R.}}:
Digital \& environment: How to evaluate server manufacturing footprint, beyond
  greenhouse gas emissions?
Boavizta
(2021).
\url{https://boavizta.org/en/blog/empreinte-de-la-fabrication-d-un-serveur}
Accessed 2025-01-20
\end{botherref}
\endbibitem

%%% 13
\bibitem[\protect\citeauthoryear{Gupta et~al.}{2022}]{10.1145/3470496.3527408}
\begin{bchapter}
\bauthor{\bsnm{Gupta}, \binits{U.}},
\bauthor{\bsnm{Elgamal}, \binits{M.}},
\bauthor{\bsnm{Hills}, \binits{G.}},
\bauthor{\bsnm{Wei}, \binits{G.-Y.}},
\bauthor{\bsnm{Lee}, \binits{H.-H.S.}},
\bauthor{\bsnm{Brooks}, \binits{D.}},
\bauthor{\bsnm{Wu}, \binits{C.-J.}}:
\bctitle{Act: designing sustainable computer systems with an architectural
  carbon modeling tool}.
In: \bbtitle{Proceedings of the 49th Annual International Symposium on Computer
  Architecture}.
\bsertitle{ISCA '22},
pp. \bfpage{784}--\blpage{799}.
\bpublisher{Association for Computing Machinery},
\blocation{New York, NY, USA}
(\byear{2022}).
\doiurl{10.1145/3470496.3527408}
\end{bchapter}
\endbibitem

%%% 14
\bibitem[\protect\citeauthoryear{Tannu and
  Nair}{2023}]{10.1145/3630614.3630616}
\begin{barticle}
\bauthor{\bsnm{Tannu}, \binits{S.}},
\bauthor{\bsnm{Nair}, \binits{P.J.}}:
\batitle{The dirty secret of ssds: Embodied carbon}.
\bjtitle{SIGENERGY Energy Inform. Rev.}
\bvolume{3}(\bissue{3}),
\bfpage{4}--\blpage{9}
(\byear{2023})
\doiurl{10.1145/3630614.3630616}
\end{barticle}
\endbibitem

%%% 15
\bibitem[\protect\citeauthoryear{Garcia~Bardon et~al.}{2020}]{9372004}
\begin{bchapter}
\bauthor{\bsnm{Garcia~Bardon}, \binits{M.}},
\bauthor{\bsnm{Wuytens}, \binits{P.}},
\bauthor{\bsnm{Ragnarsson}, \binits{L.-A.}},
\bauthor{\bsnm{Mirabelli}, \binits{G.}},
\bauthor{\bsnm{Jang}, \binits{D.}},
\bauthor{\bsnm{Willems}, \binits{G.}},
\bauthor{\bsnm{Mallik}, \binits{A.}},
\bauthor{\bsnm{Spessot}, \binits{A.}},
\bauthor{\bsnm{Ryckaert}, \binits{J.}},
\bauthor{\bsnm{Parvais}, \binits{B.}}:
\bctitle{Dtco including sustainability:
  Power-performance-area-cost-environmental score (ppace) analysis for logic
  technologies}.
In: \bbtitle{2020 IEEE International Electron Devices Meeting (IEDM)},
pp. \bfpage{41}--\blpage{414144}
(\byear{2020}).
\doiurl{10.1109/IEDM13553.2020.9372004}
\end{bchapter}
\endbibitem

%%% 16
\bibitem[\protect\citeauthoryear{Shilov}{2019}]{AnandTech2019}
\begin{botherref}
\oauthor{\bsnm{Shilov}, \binits{A.}}:
{SK Hynix Details DDR5-6400}
(2019).
\url{https://www.anandtech.com/show/13999/sk-hynix-details-its-ddr56400-dram-chip}
Accessed 2025-01-20
\end{botherref}
\endbibitem

%%% 17
\bibitem[\protect\citeauthoryear{Britton}{2023}]{Britton2023}
\begin{botherref}
\oauthor{\bsnm{Britton}, \binits{D.}}:
{ARM for WLCG}
(2023).
\url{https://indico.cern.ch/event/1289243/contributions/5583079/attachments/2735189/4756143/231016-HEPIX.pptx}
Accessed 2025-01-21
\end{botherref}
\endbibitem

%%% 18
\bibitem[\protect\citeauthoryear{Kennedy}{2019}]{amdepyc2019}
\begin{botherref}
\oauthor{\bsnm{Kennedy}, \binits{P.}}:
{AMD EPYC 7742 Benchmarks and Review Simply Peerless}
(2019).
\url{https://www.servethehome.com/amd-epyc-7742-benchmarks-and-review-simply-peerless/}
Accessed 2025-01-20
\end{botherref}
\endbibitem

%%% 19
\bibitem[\protect\citeauthoryear{{Standard Performance Evaluation
  Corporation}}{2019}]{speccpu2020}
\begin{botherref}
\oauthor{\bsnm{{Standard Performance Evaluation Corporation}}}:
{SPEC CPU 2017 Integer Speed Result}
(2019).
\url{http://spec.org/cpu2017/results/res2020q2/cpu2017-20200413-21925.pdf}
Accessed 2025-01-20
\end{botherref}
\endbibitem

%%% 20
\bibitem[\protect\citeauthoryear{Schulz}{2024}]{cernITsust}
\begin{botherref}
\oauthor{\bsnm{Schulz}, \binits{M.}}:
{ CERN-IT Sustainability}
(2024).
\url{https://indico.cern.ch/event/1450885/contributions/6251301/attachments/2984942/5257475/Sustainability-CERN-IT-TO.pdf}
Accessed 2025-06-04
\end{botherref}
\endbibitem

%%% 21
\bibitem[\protect\citeauthoryear{Zimmermann et~al.}{2012}]{ZIMMERMANN2012237}
\begin{barticle}
\bauthor{\bsnm{Zimmermann}, \binits{S.}},
\bauthor{\bsnm{Meijer}, \binits{I.}},
\bauthor{\bsnm{Tiwari}, \binits{M.K.}},
\bauthor{\bsnm{Paredes}, \binits{S.}},
\bauthor{\bsnm{Michel}, \binits{B.}},
\bauthor{\bsnm{Poulikakos}, \binits{D.}}:
\batitle{Aquasar: A hot water cooled data center with direct energy reuse}.
\bjtitle{Energy}
\bvolume{43}(\bissue{1}),
\bfpage{237}--\blpage{245}
(\byear{2012})
\doiurl{10.1016/j.energy.2012.04.037} .
\bcomment{2nd International Meeting on Cleaner Combustion (CM0901-Detailed
  Chemical Models for Cleaner Combustion)}
\end{barticle}
\endbibitem

%%% 22
\bibitem[\protect\citeauthoryear{Goda}{2021}]{electronics10243156}
\begin{botherref}
\oauthor{\bsnm{Goda}, \binits{A.}}:
Recent progress on 3d nand flash technologies.
Electronics
\textbf{10}(24)
(2021)
\doiurl{10.3390/electronics10243156}
\end{botherref}
\endbibitem

%%% 23
\bibitem[\protect\citeauthoryear{Hilty}{2015}]{zora110766}
\begin{bchapter}
\bauthor{\bsnm{Hilty}, \binits{L.}}:
\bctitle{Computing efficiency, sufficiency, and self-sufficiency: A model for
  sustainability?}
In: \bbtitle{LIMITS 2015, First Workshop on Computing Within Limits}.
\bpublisher{s.n.}, \blocation{???}
(\byear{2015}).
\burl{https://doi.org/10.5167/uzh-110766}
\end{bchapter}
\endbibitem

\end{thebibliography}
\end{document}